\documentclass[reprint, superscriptaddress, amsmath,amssymb, aps, prresearch, longbibliography, floatfix]{revtex4-2}

\usepackage{amsmath}
\usepackage{graphicx}
\usepackage{ulem}
\usepackage{bm}
\usepackage{natbib}
\usepackage{dsfont}
\usepackage{tikz}
\usepackage{epsfig}
\usepackage{feynmf}
\usepackage{blindtext, rotating}
\usepackage{mathtools}
\usepackage{dsfont}
\usepackage{subcaption}
\usepackage{physics}
\usepackage{amsfonts}
\usepackage{xcolor}
\usepackage{ragged2e}
\usepackage{siunitx}
\usepackage{comment}
\usepackage{soul}
\usepackage{lipsum}
\usepackage{hyperref} 
\hypersetup{breaklinks=true, colorlinks=true, citecolor=blue, linkcolor=cyan, urlcolor=blue,filecolor=blue}

\usepackage{xcolor} 

\DeclareCaptionJustification{justified}{\justifying}

\DeclareSIUnit{\rad}{rad}

\definecolor{bright_blue}{HTML}{85C1E9}
\definecolor{middle_blue}{HTML}{2E86C1}
\definecolor{dark_blue}{HTML}{1B4F72}

\begin{document}

\title{Passive polarization and phase stabilization scheme for Twin-Field QKD}

\author{Christiano M. S. Nascimento}
\affiliation{NITeQ, Department of Electrical Engineering, Pontifical Catholic University of Rio de Janeiro, 22451-900 Rio de Janeiro, RJ, Brazil}
\affiliation{QuIIN - Quantum Industrial Innovation, EMBRAPII CIMATEC Competence Center in Quantum Technologies, SENAI CIMATEC, Av. Orlando Gomes 1845, 41650-010, Salvador, BA, Brazil. }

\author{Felipe Calliari}
\affiliation{NITeQ, Department of Electrical Engineering, Pontifical Catholic University of Rio de Janeiro, 22451-900 Rio de Janeiro, RJ, Brazil}

\author{Guilherme P. Temporão}
\affiliation{NITeQ, Department of Electrical Engineering, Pontifical Catholic University of Rio de Janeiro, 22451-900 Rio de Janeiro, RJ, Brazil}

\begin{abstract}
Twin-Field Quantum Key Distribution requires first-order interference between coherent states sent by Alice and Bob in a mid-station Charlie. In order to obtain stable operation and maximum interferometric visibility, not only phase stabilization but also polarization control is required, especially in optical-fiber setups. In this paper, we propose an experimental setup that simultaneously provides passive stabilization of phase and polarization fluctuations by combining a Sagnac-like interferometer with a "Plug-and-Play" configuration employing Faraday mirrors. Experimental results show a net interferometric visibility maintained around 95.3\% during 72 hours of continuous operation, with a standard deviation of $0.47\%$. The setup can be straightforwardly adapted to a multi-user scenario employing either a star network, a bus topology, or a combination of both. 
\end{abstract}

\maketitle

\begin{figure*}[t]
    \centering
    \includegraphics[width=\textwidth]{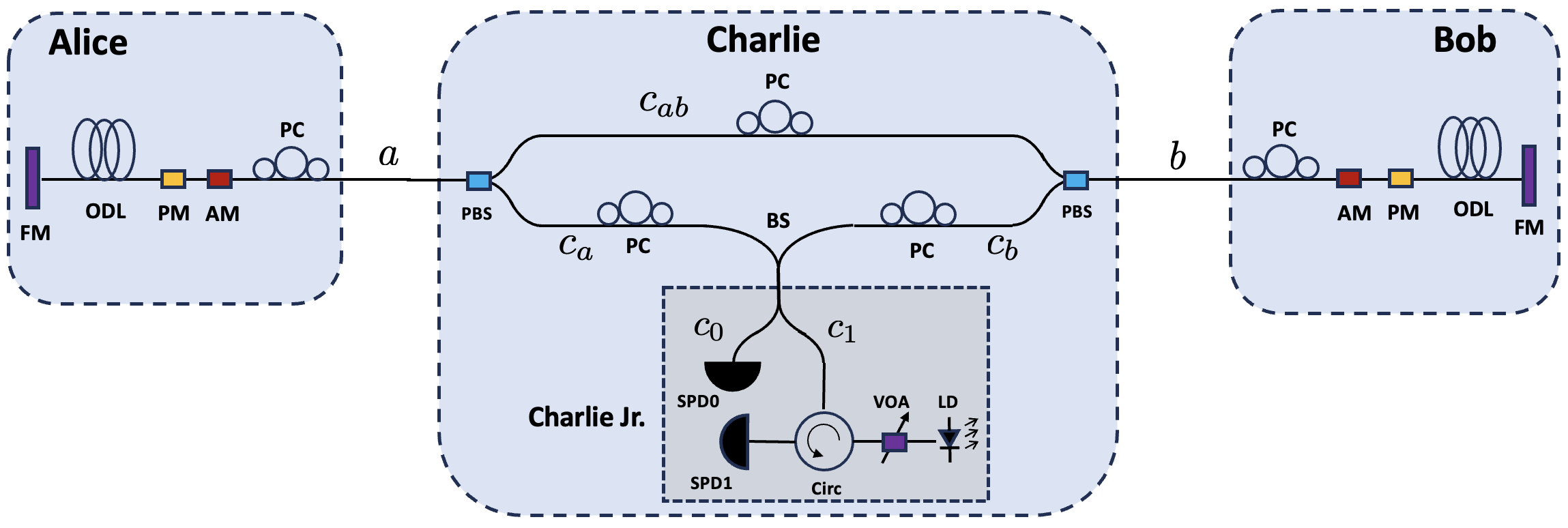}
      \caption{Basic three-node topology. Charlie prepares laser pulses that are equally split into two halves by a beamsplitter (BS), one going towards Alice (mode $c_a$) and the other towards Bob (mode $c_b$). Regardless of the polarization fluctuations introduced in modes $a$ and $b$, the two halves always recombine in the BS in the same polarization state. LD: Laser Diode; Circ: circulator; SPD: single-photon detector; PC: Fixed Polarization Controller; PBS: Polarizing Beamsplitter; PM: Phase Modulator; AM: Amplitude Modulator; FM: Faraday Mirror; ODL: Optical Delay Line; VOA: Variable Optical Attenuator.}
      \label{fig:three_node}
  \end{figure*}

\section{Introduction}

Quantum Key Distribution (QKD) provides a solution to the key distribution problem, i.e., the generation of secret random bits between two geographically distant parties, Alice and Bob, where its security is based on the laws of quantum physics \cite{Gisin:2002,Thew:2007}. However, early prepare-and-measure QKD protocols, such as BB84 \cite{Bennett:2014} or SARG04 \cite{Scarani:2004}, do not include in their security proofs all possible experimental imperfections in the devices that could be exploited by an eavesdropper, Eve. In fact, many attacks have been proposed and experimentally demonstrated, especially to single-photon detectors (SPD) \cite{Makarov:2005,Makarov:2008,Makarov:2009,Lydersen:2010,Liu:2014,Sauge:2011,Weier:2011,Qi:2007}, and a series of countermeasures to these attacks have been developed \cite{Yuan:2010,FerreiradaSilva:2012,FerreiradaSilva:2014,Lim:2015,Qian:2019,Acheva:2023}. Measurement Device Independent QKD (MDI-QKD) \cite{Lo:2012} has been proposed as a protocol that includes in its security proof the detector imperfections, which means that they can be treated in practice as black boxes. In this protocol, Alice and Bob send single photons to a mid-station Charlie, which performs a Bell state measurement based on two-photon interference in a beamsplitter.

Twin-Field (TF) QKD has been proposed as an alternative to the original MDI-QKD protocol that does not require two-photon interference; instead, it employs "classical" (first-order) interference of coherent states in a beamsplitter \cite{Lucamarini:2018,Yin:2019}. The main advantage is the robustness to losses: if the total channel transmission - considering the Alice-Charlie and Bob-Charlie links - is given by $t$, then the raw key rate is proportional to $\sqrt{t}$, thus surpassing the PLOB bound \cite{Pirandola:2017}. However, differently from standard MDI-QKD, TF-QKD requires a distributed phase reference: assuming that the coherent states at Charlie's beamsplitter inputs, arriving from Alice and Bob, are described by $\ket{\sqrt{\mu_A}e^{j\alpha}}$ and $\ket{\sqrt{\mu_B}e^{j\beta}}$, the phase difference $\alpha-\beta$ must be kept stable over time. Therefore, different solutions for this problem have been proposed in recent experimental implementations \cite{Chen:2021,Liu:2021,Wang:2022,Yang:2023,Yuan:2022,Wei:2023,Zhou:2023}. One of these solutions employs a clever arrangement of Alice, Bob and Charlie along a Sagnac interferometer, which provides passive compensation for fluctuations in the optical path length \cite{Zhong:2019,Zhong:2022}. Despite its inherent shortcomings, such as increased susceptibility to Rayleigh backscattering noise \cite{Mandil:2025}, this solution has been successfully implemented. It is important to point out, however, that it is sensitive to random polarization fluctuations along the optical fiber: if the two pulses propagating back to Charlie do not have matching polarization states, the interferometric visibility will be reduced as $V = \sqrt{\mathcal{F}}$, where $\mathcal{F} = |\langle \psi|\phi\rangle |^2$ is the fidelity between the two polarization states $\ket{\psi},\ket{\phi}$. It is required, therefore, to implement a polarization control system. There are many possible solutions to this problem in the literature \cite{Xavier:2011a,Xavier:2011b,Xavier:08,Faria:2008,Ramos:2022}, but up to this moment, all of them involve some kind of active control, which increases the complexity of the implementation, or adjustments over time, which reduce the net key rate.

In this paper, we introduce a method to passively compensate polarization fluctuations, whereas keeping the passive phase compensation inherent to the Sagnac implementation. The method is directly based on the "Plug and Play" setup proposed by Ribordy et al \cite{Ribordy:1998} in the context of standard prepare-and-measure QKD. By employing Faraday mirrors (FM) and polarizing beamsplitters (PBS) in a specific arrangement, we show that a Sagnac-like interferometer is obtained, thus keeping the ability to compensate phase fluctuations and adding passive polarization stabilization. Differently from the original Sagnac proposal, which uses a ring topology, this work uses star or bus topologies (or a combination of both), where Charlie acts as the central node. It should be mentioned that a "Plug and Play" arrangement for TF-QKD has already been proposed \cite{Xue:2021}, but it lacks the passive phase stabilization property of this work.

In section \ref{sec:three_node}, the standard setup for a three-node network is presented, whereas section \ref{sec:Multi-node} shows a generalization for multiple nodes. Section \ref{sec:Experimental_setup} contains the experimental results, showing that the configuration is indeed stable to polarization and phase fluctuations; sections \ref{sec:discussion} and \ref{sec:conclusions} discuss the potential advantages of this proposal and draw the conclusions. 

\section{Basic Three-Node Topology}\label{sec:three_node}

The standard configuration for a three-node network - connecting Alice, Bob and Charlie - is represented in Fig. 1. Similarly to a Sagnac configuration, a faint laser pulse in an arbitrary polarization state $\ket{\textbf{SOP}}$ is split into a "clockwise" and a "counterclockwise" component (in analogy to a Sagnac interferometer)  by a beamsplitter (BS): the first going towards Alice in mode $c_a$, the second towards Bob in mode $c_b$. Fixed polarization controllers (PC) ensure that the polarization states at the PBS always match one of its eigenstates - namely, the one which is transmitted, which we will assume is the horizontal polarization $\ket{H}$. This means that the pulses traveling from the output of Charlie Jr. to the outputs of Charlie are described by the following sequence of operations:
\begin{align}
    \label{eq:1}
        \ket{\psi}_0 &= \ket{c_1}\otimes \ket{\textbf{SOP}}\nonumber \\
        &\xrightarrow{\text{BS}} \frac{1}{\sqrt{2}}\left(i\ket{c_a,\textbf{-k}}+\ket{c_b,\textbf{+k}}\right)\otimes \ket{\textbf{SOP}}\nonumber \\ 
        &\xrightarrow{\text{PC, PBS}} \frac{1}{\sqrt{2}}\left(i\ket{a,\textbf{-k}}+\ket{b,\textbf{+k}}\right)\otimes \ket{H},
\end{align}
where $\pm \textbf{k}$ indicates the direction of propagation, which we assume is positive from the left to the right, and the polarization controllers (PC) introduce a unitary operation $U_{PC}$ such that $U_{PC}\ket{\textbf{SOP}}=\ket{H}$. Note that even though we are using quantum mechanics notation, the pulses are entirely classical. 

Due to the presence of a Faraday mirror (FM) in each endpoint - Alice's and Bob's offices - the polarization state of the optical pulses that return to Charlie always correspond to the horizontal state $\ket{V}$, irrespective of the unitary polarization transformation introduced by the optical fibers in modes $a$ and $b$. The returning pulses are now described by:
\begin{equation}
    \label{eq:2}
        \ket{\psi}_1 = \frac{1}{\sqrt{2}}\left(i\ket{a,\textbf{+k}}+e^{i\phi}\ket{b,\textbf{-k}}\right)\otimes \ket{V},
\end{equation} 
where $\phi$ is a random relative phase. Now, both pulses leave their respective PBS from the upper ports - Alice's pulse going towards Bob and vice-versa through mode $c_{ab}$. Another fixed PC ensures that the PBS eigenstates match each other. Now, both pulses are reflected by their respective PBS, resulting in the pulse that was in mode $a$ now going towards Bob in mode $b$ and vice-versa:
\begin{equation}
    \label{eq:3}
        \ket{\psi}_2 = \frac{1}{\sqrt{2}}\left(i\ket{b,\textbf{+k}}+e^{i\phi_1}\ket{a,\textbf{-k}}\right)\otimes \ket{V},
\end{equation} 
where the relative phase is unchanged as both pulse halves have propagated through the same optical path. The same effect from the previous stage will happen: as both pulses were originally in the $\ket{V}$ state, the FMs will act such that they return to the original $\ket{H}$ state, such that the PBSs forward them to the BS and to Charlie Jr. following the sequence:
\begin{align}
    \label{eq:4}
        \ket{\psi}_3 &= \frac{1}{\sqrt{2}}\left(ie^{i\phi}\ket{c_b,\textbf{-k}}+e^{i\phi}\ket{c_a,\textbf{+k}}\right)\otimes \ket{H}\nonumber \\
        &\xrightarrow{\text{PBS, PC}} \frac{1}{\sqrt{2}}\left(i\ket{c_b,\textbf{-k}}+\ket{c_a,\textbf{+k}}\right)\otimes\ket{\textbf{SOP}}\nonumber\\
        &\xrightarrow{\text{BS}}\ket{c_1}\otimes\ket{\textbf{SOP}},
\end{align} 
where one can notice that the term $e^{i\phi}$ in the first line is now a global phase and can be neglected. Again, $\ket{\textbf{SOP}} = U_{PC}^\dagger\ket{H}$. In other words, when impinging on the BS, there will be destructive interference in the lower left port and constructive interference in the lower right port, meaning that all light will return to the Circulator and be forwarded to the single-photon detector SPD1.

It is straightforward to notice that, if nothing happens to the optical modes $a$ and $b$ during the pulse propagation time, the optical paths taken by the two pulses are identical - in fact, they are the same. That's why the relative phase added in the last stage was also $\phi$. In other words, any path length fluctuation acting on a time scale longer than the round-trip time will be experienced by both pulses and cancel out, just as in a standard Sagnac interferometer. The same limitations take place here: the longer the interferometer - i.e., Alice-Bob distance - the less resilient to fast phase fluctuations the setup becomes.

It is also important to point out that the amplitude and phase modulation stages by Alice and Bob are not carried out until the last moment. For example, Alice's Amplitude Modulator (AM) and Phase Modulator (PM) are not activated until the original "counterclockwise" pulse passes through Bob and Charlie. It is only at this point, where the pulses become ready to go back to Charlie towards the BS, that Alice and Bob apply their AM/PM pulses. Moreover, until this very last moment, the pulses are not at the single-photon level, as they do not contain any information. This ensures that the only relevant attenuation coefficients correspond to the one-way propagation from Alice to Charlie and Bob to Charlie. A side effect of this feature is the presence of unwanted Rayleigh backscattered photons that go towards Charlie's detectors, similarly to what has been measured in the standard Sagnac configuration \cite{Mandil:2025}.

It should be mentioned that the PCs present in Alice and Bob's offices are required not because of polarization control, but because the modulators usually have preferred polarizations due to polarization dependent loss (PDL). In some cases, an additional PC between the AM and PM may be needed for this reason. Moreover, the optical delay lines (ODL) in Alice and Bob are required to avoid the coexistence of pulses traveling in opposite directions in the modulators.

\section{Multi-Node Topology}\label{sec:Multi-node}

It turns out that the basic scheme presented in the previous section can be generalized to a network with an arbitrary number of users, as shown in Fig. \label{fig:many_node}. Here we show a mixed network topology: starting from a star network, where Charlie has 4 optical fibers connected to his office. Each branch defined by each fiber can actually be comprised of multiple users in tandem, as in a bus topology. This is possible thanks to the FM inside Charlie's station. In this example, we only show two users in such a disposition, Emily and Frank; but, in principle, the same pattern can be repeated indefinitely (with $N$ users in tandem in each branch). The number of branches can also be increased by appropriately increasing the number of ports of the switches. 

The main idea is simple: once two users agree on generating a secret key, the optical switches (SW) are configured in a specific way, and they remain in the same configuration until the protocol is finished. This means that the switches do not need to be high-speed; on the contrary, slow (and low-loss) switches can be deployed.

Once the switches have been configured, the protocol works in the exact same way as in the previous section. Tab. \ref{Tab:First} illustrates the switch configurations for establishing a connection between Alice-Bob, Debbie-Emily and Emily-Frank. Note that switch $SW_A$ is 1x2, whereas switch $SW_B$ is 2x4, so in this case there are two columns, each one showing which external port ({\it a,b,d,e}) is connected to each internal port (1, 2).

\begin{figure}[t!]
    \centering
    \includegraphics[width=\columnwidth]{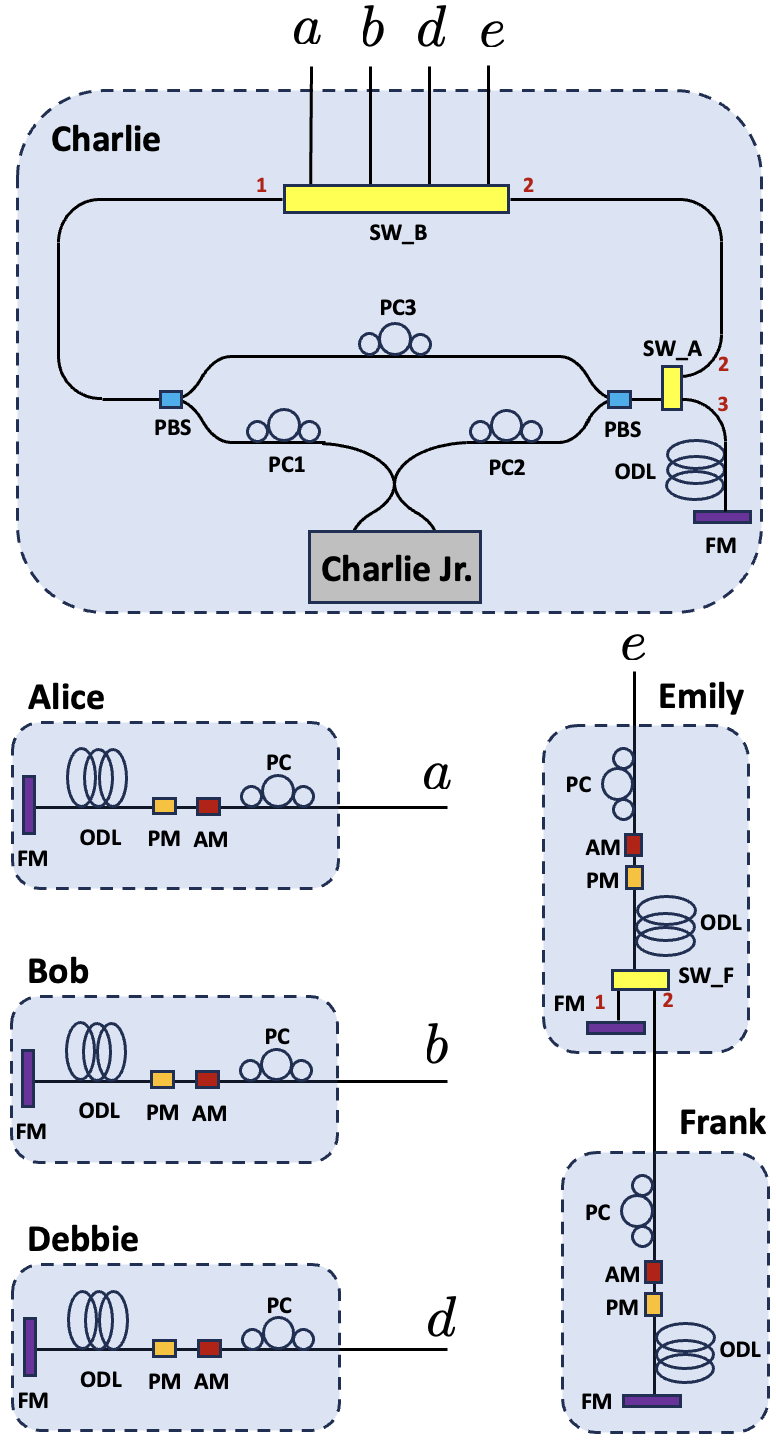}
      \caption{Topology for an arbitrary number of nodes. In this example, there are 5 users connected to Charlie: Alice, Bob, Debbie, Emily and Frank, interconnected in a hybrid topology. Charlie acts as the central node, spanning a star-shaped network; however, each branch can connect any number of users in tandem, which is similar to a bus topology. Any user in this network can exchange keys with any other user by properly configuring the optical switches (SW). "Charlie Jr." contains the same components as in Fig. 1.}
      \label{fig:many_node}
  \end{figure}

\begin{table}[t]
    \caption{\label{tab:one}Examples of switch configurations for a few connections. The symbol "-" indicates the configuration is irrelevant for that particular connection.}
    \label{Tab:First}
    \begin{ruledtabular}
    \begin{tabular}{ccccccc}
        Connection & $SW_A$  & $SW_B(1)$ & $SW_B(2)$ & $SW_F$  \\
        \hline
        Alice-Bob& 2 & $a$ & $b$ & -  \\
        Debbie-Emily & 2 & $d$ & $e$ & -  \\
        Emily-Frank & 3 & $e$ & - & 2 \\
    \end{tabular}
    \end{ruledtabular}
\end{table}
The configurations in Tab. \ref{Tab:First} are only examples and are not unique, i.e., the connection between two users can usually be done in two different ways, corresponding to swapping their positions. This redundancy is needed for allowing every user to be able to connect to every other user.

Note the presence in Fig. \label{fig:many_node} of an optical delay line (ODL) on the upper path inside Charlie and a FM connected to Switch A. These components are necessary for connections between users in the same branch - Emily and Frank in this example. The incoming pulse towards the right in this case is directly reflected by the FM and needs to be delayed to avoid overlap with the pulse originally going towards the left. 

For adding more users in the same branch, one should just add more copies of Emily in tandem. The number of users that can be connected this way is only limited by the interferometer size, as discussed before. 

Finally, it should also be mentioned that this setup is readily compatible with hybrid quantum networks where fiber-optical and free-space channels coexist. Note that, in Fig. \label{fig:many_node}, any of the users Alice, Bob or Debbie could employ a free-space link for implementation of optical modes $a$, $b$ or $d$. In a standard Sagnac loop topology, the addition of a free-space link would impact all users, whereas in our proposal the free-space channel is used only while the corresponding user is generating a secret key with one of the other parties.

\section{Experimental Setup}\label{sec:Experimental_setup}

\begin{figure}[!h]
    \centering
    \includegraphics[scale=0.45]{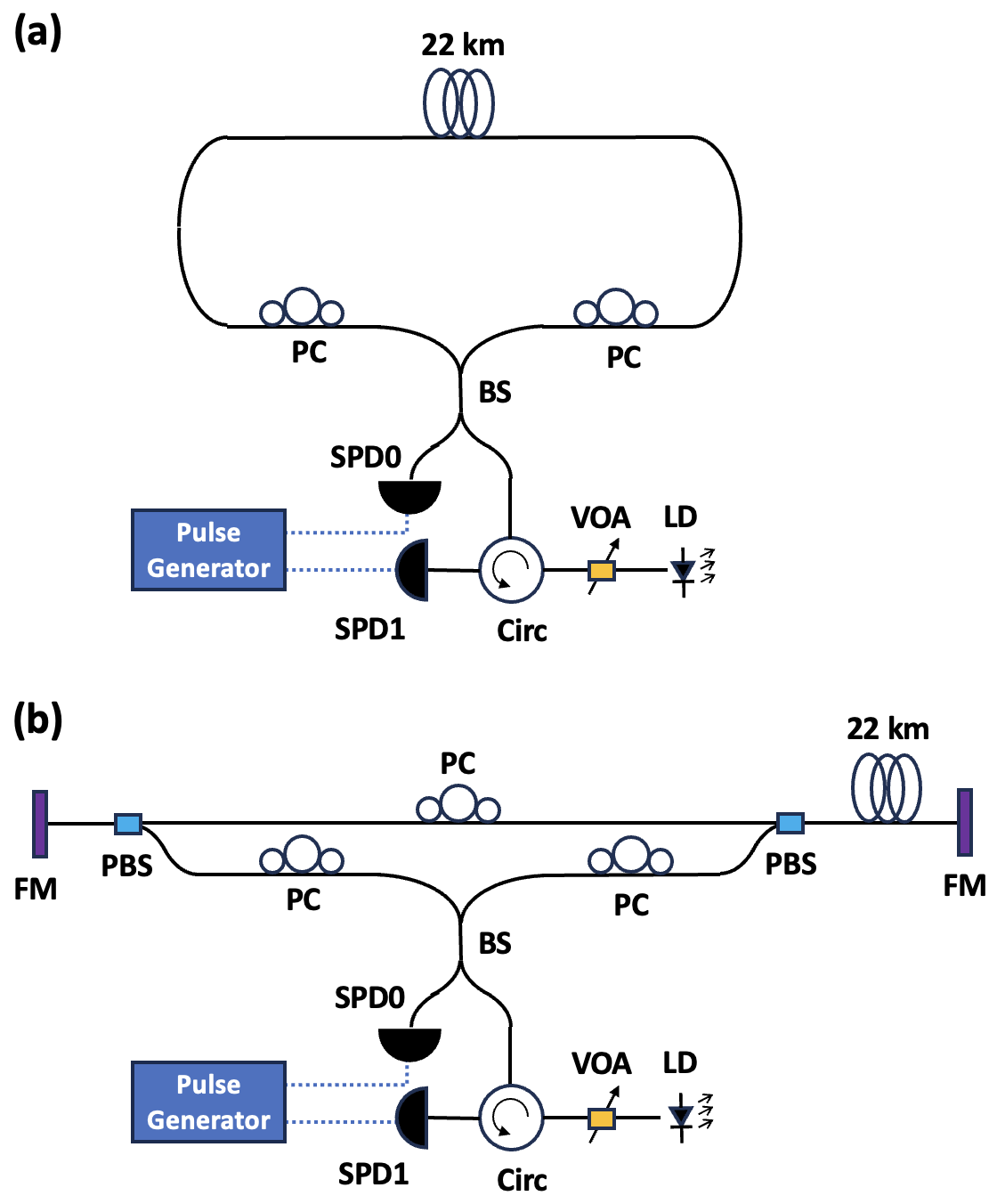}
    \caption{Experimental setups. (a) Standard Sagnac inteferometer; (b) Modifier Sagnac interferometer employing Faraday Mirrors. In both cases a 22-km fiber spool is employed.}
    \label{fig:Experimental setup}
\end{figure}

To experimentally demonstrate simultaneous polarization and phase stability, we implemented our scheme and compared the results with a standard Sagnac interferometer, as shown in Fig. \ref{fig:Experimental setup}. In both cases, we employed an attenuated telecommunication laser diode (LD) at 1550nm, operated at a continuous wave (CW) regime, i.e., no modulators were implemented. Therefore, in the ideal case, all light should return to the optical circulator.

To simulate the optical fiber link between the three nodes, a 22 km standard single-mode fiber spool was used. A pulse generator was employed as an external trigger to both InGaAs avalanche single-photon detection modules (SPD), with a gating rate of 2MHz and 20ns pulse width. Both SPDs were configured to an efficiency of 10\% and deadtime of 1 $\mu$s, resulting in dark count rates of about 153 and 244 Hz for SPD0 and SPD1, respectively. 

In both experiments, we adjusted the PCs such that the maximum possible interferometric visibility was achieved. Moreover, the variable optical attenuator (VOA) was adjusted such that a maximum count rate of about 7kHz was obtained in detector SPD1. Then we accumulated photon-counting statistics over a time period of 72 hours such that the effects of slow time-varying polarization fluctuations in the optical fiber spool could be observed. We used an acquisition rate of 1 Hz and afterwards we employed a moving-average filter of 10 samples, emulating an integration time of 10 seconds.

\begin{figure}[htpb]
    \centering
    \includegraphics[scale=0.255]{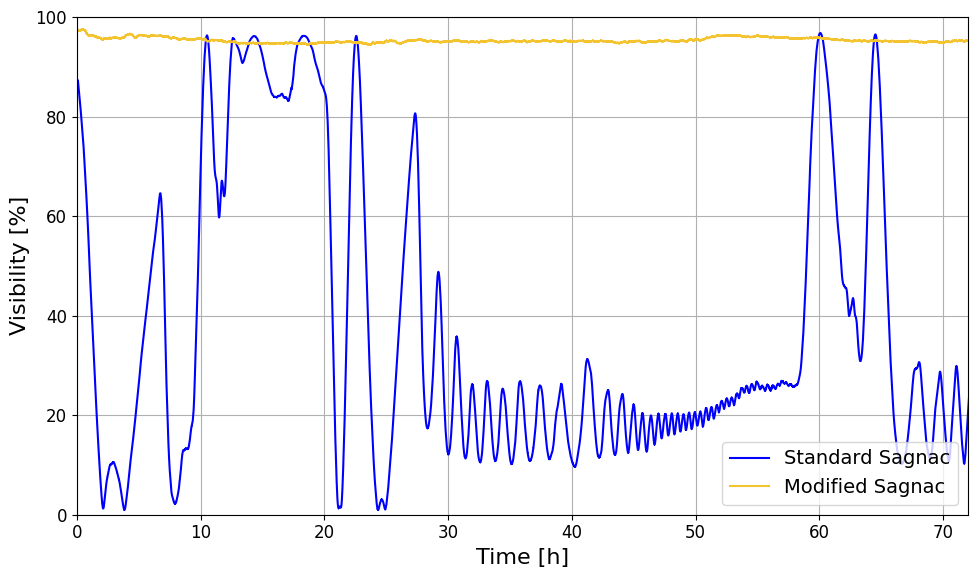}
    \caption{Net visibility of the interferometers over 72 hours of continuous operation. The blue curve depicts the standard Sagnac interferometer, whereas the yellow curve depicts the modified Sagnac interferometer (proposed scheme).}
    \label{fig:results_80_hrs}
\end{figure}

As depicted in Fig. \ref{fig:results_80_hrs} and in Tab. \ref{tab:Tab_two}, the visibility of the standard Sagnac interferometer heavily changed over time, from approximately 97\% to 1\%, demonstrating that the polarization state varies significantly over time. This is due to variations in the mechanical properties of the optical fiber, such as vibration and temperature. However, the visibility of the new scheme barely changed, from approximately 94\% to 98\%, which shows that polarization is not an issue and can be handled with passive optical components. It should be mentioned that these values correspond to net visibilities, i.e., neglecting the detector dark counts.

It is worth noting that both setups were implemented under laboratory conditions, which means that, for metropolitan fiber scenarios, polarization is expected to be more unstable. 

\begin{table}[htpb]
    \caption{\label{tab:Tab_two}Statistical analysis of net visibilities for both Sagnac setups (\%)}
    \begin{ruledtabular}
    \begin{tabular}{lcccc}
        Setup & Maximum & Minimum & Average & \begin{tabular}[c]{@{}c@{}}Standard\\deviation\end{tabular} \\
        \hline
        Standard Sagnac & 96.8 & 0.933 & 35.9 & 27.7 \\
        Modified Sagnac & 97.6 & 94.4 & 95.3 & 0.476 \\
    \end{tabular}
    \end{ruledtabular}
\end{table}

\section{Discussion}\label{sec:discussion}

In the previous section, the interferometric visibilities were measured in an experimental configuration similar to a QKD scenario, where there is an average of about one photon per "pulse" (detection window) in the output of the VOA. However, as a gate width of 20 ns was used, the detector dark counts were higher than they would in a QKD scenario, where typically windows of 1 ns are employed. This explains why the dark counts were not negligible and had to be subtracted in order to obtain the net visibility. The raw visibility of the modified Sagnac setup was 91.4\%, with a standard deviation of 0.33\%.

Moreover, as previously mentioned, Rayleigh Backscattering plays an important role in the visibility \cite{Mandil:2025}. We also noticed that the PBS extinction ratio, i.e., its finite PDL \cite{Calliari:2019}, could be a limiting factor for the visibility. We address these two points below.

\subsection{\label{sec:rayleigh}Rayleigh Backscattering}

As our interferometer operates in CW regime, all backscattered photons, irrespective of their arrival times, will incoherently add to the signal photons that follow the intended optical path.

An infinitesimal contribution of an element of length $dz$ for Rayleigh backscattering is given by:
\begin{equation}
    \label{eq:rayleigh1}
dP_R(z) = P_0 \cdot \gamma \cdot e^{-2\alpha' z} \, dz
\end{equation}
where $P_0$ is the launch power, $\gamma$ is the backscattering coefficient and $\alpha'$ is the fiber attenuation coefficient (in Np/km).  Integrating over the total (one-way) fiber length $L$, we get
\begin{equation}
    \label{eq:rayleigh2}
P_R = \int_0^L P_0 \cdot \gamma \cdot e^{-2\alpha' z} \, dz = 
P_0 \cdot \gamma \cdot \frac{1 - e^{-2\alpha' L}}{2\alpha'}
\end{equation}
The signal photons, on the other hand, are merely attenuated by a factor $e^{-2\alpha ' L}$. The resulting visibility is given by:
\begin{equation}
    \label{eq:rayleigh3}
V = \frac{t^2-(1-t^2)\frac{\gamma}{2\alpha '}}{t^2+(1-t^2)\frac{\gamma}{2\alpha '}}
\end{equation}
where $t = e^{-\alpha ' L}$ is the one-way channel transmission. Using $\gamma = 8\times 10^{-5}\text{km}^{-1}$ \cite{Derickson}, $\alpha' = 0.053$ (corresponding to 0.23 dB/km) and $L = 22$ km, we get $V\approx 98.4\%$. This value is compatible with our measurements, as the highest measured visibility was about 97.6\%. Note that we are not considering polarization effects here; the actual visibility would be slightly higher, because the PBS would filter out a portion of the Rayleigh backscattered light. 

\subsection{\label{sec:extinction_ratio}Polarization Misalignment}

The use of manual PCs introduces a small misalignment between the input polarization states in the PBSs and their eigenstates. Let $P_0$ be the output laser power impinging in Charlie's BS. Consider now the PCs in modes $c_a$, $c_b$ produce polarization states that are misaligned with the PBSs, such that the optical powers in the horizontal and vertical polarizations are given by $P_H = \xi P_0$ and $P_V = (1-\xi) P_0$, where $\xi \equiv | \langle \psi | H \rangle |^2 $ and $\ket{\psi}$ is the polarization state immediately after the PCs. We assume the same polarization state in both modes $c_a$,$c_b$ for simplicity. The optical powers immediately $\it{after}$ the PBS in both sides, neglecting their insertion losses, are given by
\begin{equation}
    \label{eq:5}
        \frac{P}{P_0} = (1-\eta) \xi + \eta (1-\xi)
\end{equation} 
where $\eta$ is the fraction of unwanted (vertical) polarization that couples into the spatial mode that ideally corresponds to horizontal polarization. Note that this is a partially polarized state, which means that, after reflection in the FM, a fraction of the power will couple back into modes $c_a$,$c_b$. As the polarization state which experienced a loss $\eta$ in the outward direction will now undergo a $1-\eta$ loss when coming back towards the PBSs, the optical powers in these modes, which we can classify as noise, are given by:
\begin{equation}
    \label{eq:6}
        \frac{P_{noise}}{P_0} = \eta(1-\eta)(1-\xi)
\end{equation} 
Assuming that the coherence length of the laser is much shorter than the total optical path, this spurious power adds incoherently with the legitimate power that follows the full optical path, which is given by
\begin{equation}
    \label{eq:7}
 \frac{P_{signal}}{P_0} \approx t^2\xi^2(1-\eta)^4 
\end{equation} 
where $t$ is the one-way channel transmission as before. Note that there is an additional $\xi$ factor corresponding to a polarization misalignment in the upper path $c_{ab}$. This results in a visibility given by:
\begin{equation}
    \label{eq:8}
        V \approx \frac{t^2\xi^2-\eta(1-\xi)}{t^2\xi^2+\eta(1-\xi)}
\end{equation} 
where the approximation holds only if $\eta \ll 1$. The extinction ratio of a PBS is given by $(1-\eta)/\eta$, expressed in decibels. In our experiment, both PBSs have an extinction ratio of about 18dB, which corresponds to $\eta \approx 0.0156$. From the behavior of the single photon detection counts over time, we estimate an alignment factor $\xi \approx 0.97$, i.e., a polarization misalignment of 3\% in each PBS.

Moreover, we measured an attenuation of 11.3 dB in the fiber spools (including insertion losses of all components), corresponding to $t \approx 0.074$. Plugging these values into Eq. \ref{eq:8}, we get $V \approx 98.7\%$, which means that, depending on the extinction ratio of the PBS units, the polarization misalignment issue can be comparable with Rayleigh backscattering. A possible way to circumvent this effect is the deployment of polarization-maintaining (PM) fibers in modes $c_a$, $c_b$.

It should also be noted that this visibility would be higher in an actual use of the interferometer for TF-QKD. In the pulsed regime, the spurious photons from any residual misalignment would be filtered by properly gating the detectors, as the noise photons arrive much earlier than the signal ones. This is also required for filtering Rayleigh backscattered light \cite{Mandil:2025}.\\

\section{Conclusions}\label{sec:conclusions}

We have presented a modification to the Sagnac interferometer that passively stabilizes power fluctuations caused by random polarization rotations. By incorporating a pair of Faraday mirrors, we experimentally demonstrate that a net interferometric visibility of 95.3\% that can be sustained over days without any active polarization control — something unachievable in standard Sagnac configurations. In our implementation, the visibility was primarily limited by both Rayleigh backscattering and a slight polarization misalignment combined with a poor extinction ratio of the polarizing beamsplitters used, both of which can be readily improved, e.g. by using polarization maintaining fibers between the beamsplitter and the polarizing beamsplitters. Nonetheless, irrespective of the visibility value, we demonstrated a very stable operation, with a standard deviation of only $0.476\%$ during an operation time of 72 hours under conditions where a standard Sagnac configuration showed an average visibility of about 36\%.

Even though our proposal can be applied to any Sagnac interferometer, it is particularly well-suited for Twin-Field Quantum Key Distribution, as it simultaneously addresses two major challenges: phase and polarization fluctuations along the optical path. Furthermore, we show that the setup can be easily extended to a QKD network with multiple users, supporting at least three distinct topologies—star, bus, and mixed configurations. Moreover, it has the inherent advantage of being naturally compatible with hybrid architectures combining optical fiber and free-space channels.

\section*{Acknowledgments}

The authors acknowledge financial support from FAPESP grant number 2021/06823-5 - MCTIC/CGI; and CNPq grant number 409596/2022-1.
This work has been partially funded by QuIIN - Quantum Industrial Innovation, EMBRAPII CIMATEC Competence Center in Quantum Technologies, with financial resources from the PPI IoT/Manufatura 4.0 of the MCTI grant number 053/2023, signed with EMBRAPII.

\bibliography{main}

\begin{thebibliography}{41}%
\makeatletter
\providecommand \@ifxundefined [1]{%
 \@ifx{#1\undefined}
}%
\providecommand \@ifnum [1]{%
 \ifnum #1\expandafter \@firstoftwo
 \else \expandafter \@secondoftwo
 \fi
}%
\providecommand \@ifx [1]{%
 \ifx #1\expandafter \@firstoftwo
 \else \expandafter \@secondoftwo
 \fi
}%
\providecommand \natexlab [1]{#1}%
\providecommand \enquote  [1]{``#1''}%
\providecommand \bibnamefont  [1]{#1}%
\providecommand \bibfnamefont [1]{#1}%
\providecommand \citenamefont [1]{#1}%
\providecommand \href@noop [0]{\@secondoftwo}%
\providecommand \href [0]{\begingroup \@sanitize@url \@href}%
\providecommand \@href[1]{\@@startlink{#1}\@@href}%
\providecommand \@@href[1]{\endgroup#1\@@endlink}%
\providecommand \@sanitize@url [0]{\catcode `\\12\catcode `\$12\catcode `\&12\catcode `\#12\catcode `\^12\catcode `\_12\catcode `\%12\relax}%
\providecommand \@@startlink[1]{}%
\providecommand \@@endlink[0]{}%
\providecommand \url  [0]{\begingroup\@sanitize@url \@url }%
\providecommand \@url [1]{\endgroup\@href {#1}{\urlprefix }}%
\providecommand \urlprefix  [0]{URL }%
\providecommand \Eprint [0]{\href }%
\providecommand \doibase [0]{https://doi.org/}%
\providecommand \selectlanguage [0]{\@gobble}%
\providecommand \bibinfo  [0]{\@secondoftwo}%
\providecommand \bibfield  [0]{\@secondoftwo}%
\providecommand \translation [1]{[#1]}%
\providecommand \BibitemOpen [0]{}%
\providecommand \bibitemStop [0]{}%
\providecommand \bibitemNoStop [0]{.\EOS\space}%
\providecommand \EOS [0]{\spacefactor3000\relax}%
\providecommand \BibitemShut  [1]{\csname bibitem#1\endcsname}%
\let\auto@bib@innerbib\@empty
\bibitem [{\citenamefont {Gisin}\ \emph {et~al.}(2002)\citenamefont {Gisin}, \citenamefont {Ribordy}, \citenamefont {Tittel},\ and\ \citenamefont {Zbinden}}]{Gisin:2002}%
  \BibitemOpen
  \bibfield  {author} {\bibinfo {author} {\bibfnamefont {N.}~\bibnamefont {Gisin}}, \bibinfo {author} {\bibfnamefont {G.}~\bibnamefont {Ribordy}}, \bibinfo {author} {\bibfnamefont {W.}~\bibnamefont {Tittel}},\ and\ \bibinfo {author} {\bibfnamefont {H.}~\bibnamefont {Zbinden}},\ }\bibfield  {title} {\bibinfo {title} {Quantum cryptography},\ }\href@noop {} {\bibfield  {journal} {\bibinfo  {journal} {Rev. Mod. Phys.}\ }\textbf {\bibinfo {volume} {74}},\ \bibinfo {pages} {145} (\bibinfo {year} {2002})}\BibitemShut {NoStop}%
\bibitem [{\citenamefont {Gisin}\ and\ \citenamefont {Thew}(2007)}]{Thew:2007}%
  \BibitemOpen
  \bibfield  {author} {\bibinfo {author} {\bibfnamefont {N.}~\bibnamefont {Gisin}}\ and\ \bibinfo {author} {\bibfnamefont {R.}~\bibnamefont {Thew}},\ }\bibfield  {title} {\bibinfo {title} {Quantum communication},\ }\href {https://doi.org/10.1038/nphoton.2007.22} {\bibfield  {journal} {\bibinfo  {journal} {Nature Photonics}\ }\textbf {\bibinfo {volume} {1}},\ \bibinfo {pages} {165} (\bibinfo {year} {2007})}\BibitemShut {NoStop}%
\bibitem [{\citenamefont {Bennett}\ and\ \citenamefont {Brassard}(2014)}]{Bennett:2014}%
  \BibitemOpen
  \bibfield  {author} {\bibinfo {author} {\bibfnamefont {C.~H.}\ \bibnamefont {Bennett}}\ and\ \bibinfo {author} {\bibfnamefont {G.}~\bibnamefont {Brassard}},\ }\bibfield  {title} {\bibinfo {title} {Quantum cryptography: Public key distribution and coin tossing},\ }\href {https://doi.org/https://doi.org/10.1016/j.tcs.2014.05.025} {\bibfield  {journal} {\bibinfo  {journal} {Theoretical Computer Science}\ }\textbf {\bibinfo {volume} {560}},\ \bibinfo {pages} {7} (\bibinfo {year} {2014})},\ \bibinfo {note} {theoretical Aspects of Quantum Cryptography – celebrating 30 years of BB84}\BibitemShut {NoStop}%
\bibitem [{\citenamefont {Scarani}\ \emph {et~al.}(2004)\citenamefont {Scarani}, \citenamefont {Ac\'{\i}n}, \citenamefont {Ribordy},\ and\ \citenamefont {Gisin}}]{Scarani:2004}%
  \BibitemOpen
  \bibfield  {author} {\bibinfo {author} {\bibfnamefont {V.}~\bibnamefont {Scarani}}, \bibinfo {author} {\bibfnamefont {A.}~\bibnamefont {Ac\'{\i}n}}, \bibinfo {author} {\bibfnamefont {G.}~\bibnamefont {Ribordy}},\ and\ \bibinfo {author} {\bibfnamefont {N.}~\bibnamefont {Gisin}},\ }\bibfield  {title} {\bibinfo {title} {Quantum cryptography protocols robust against photon number splitting attacks for weak laser pulse implementations},\ }\href {https://doi.org/10.1103/PhysRevLett.92.057901} {\bibfield  {journal} {\bibinfo  {journal} {Phys. Rev. Lett.}\ }\textbf {\bibinfo {volume} {92}},\ \bibinfo {pages} {057901} (\bibinfo {year} {2004})}\BibitemShut {NoStop}%
\bibitem [{\citenamefont {Makarov}\ and\ \citenamefont {Hjelme}(2005)}]{Makarov:2005}%
  \BibitemOpen
  \bibfield  {author} {\bibinfo {author} {\bibfnamefont {V.}~\bibnamefont {Makarov}}\ and\ \bibinfo {author} {\bibfnamefont {D.~R.}\ \bibnamefont {Hjelme}},\ }\bibfield  {title} {\bibinfo {title} {Faked states attack on quantum cryptosystems},\ }\href {https://doi.org/10.1080/09500340410001730986} {\bibfield  {journal} {\bibinfo  {journal} {Journal of Modern Optics}\ }\textbf {\bibinfo {volume} {52}},\ \bibinfo {pages} {691} (\bibinfo {year} {2005})},\ \Eprint {https://arxiv.org/abs/https://doi.org/10.1080/09500340410001730986} {https://doi.org/10.1080/09500340410001730986} \BibitemShut {NoStop}%
\bibitem [{\citenamefont {Makarov}\ \emph {et~al.}(2006)\citenamefont {Makarov}, \citenamefont {Anisimov},\ and\ \citenamefont {Skaar}}]{Makarov:2008}%
  \BibitemOpen
  \bibfield  {author} {\bibinfo {author} {\bibfnamefont {V.}~\bibnamefont {Makarov}}, \bibinfo {author} {\bibfnamefont {A.}~\bibnamefont {Anisimov}},\ and\ \bibinfo {author} {\bibfnamefont {J.}~\bibnamefont {Skaar}},\ }\bibfield  {title} {\bibinfo {title} {Effects of detector efficiency mismatch on security of quantum cryptosystems},\ }\href {https://doi.org/10.1103/PhysRevA.74.022313} {\bibfield  {journal} {\bibinfo  {journal} {Phys. Rev. A}\ }\textbf {\bibinfo {volume} {74}},\ \bibinfo {pages} {022313} (\bibinfo {year} {2006})}\BibitemShut {NoStop}%
\bibitem [{\citenamefont {Makarov}(2009)}]{Makarov:2009}%
  \BibitemOpen
  \bibfield  {author} {\bibinfo {author} {\bibfnamefont {V.}~\bibnamefont {Makarov}},\ }\bibfield  {title} {\bibinfo {title} {Controlling passively quenched single photon detectors by bright light},\ }\href {https://doi.org/10.1088/1367-2630/11/6/065003} {\bibfield  {journal} {\bibinfo  {journal} {New Journal of Physics}\ }\textbf {\bibinfo {volume} {11}},\ \bibinfo {pages} {065003} (\bibinfo {year} {2009})}\BibitemShut {NoStop}%
\bibitem [{\citenamefont {Lydersen}\ \emph {et~al.}(2010)\citenamefont {Lydersen}, \citenamefont {Wiechers}, \citenamefont {Wittmann}, \citenamefont {Elser}, \citenamefont {Skaar},\ and\ \citenamefont {Makarov}}]{Lydersen:2010}%
  \BibitemOpen
  \bibfield  {author} {\bibinfo {author} {\bibfnamefont {L.}~\bibnamefont {Lydersen}}, \bibinfo {author} {\bibfnamefont {C.}~\bibnamefont {Wiechers}}, \bibinfo {author} {\bibfnamefont {C.}~\bibnamefont {Wittmann}}, \bibinfo {author} {\bibfnamefont {D.}~\bibnamefont {Elser}}, \bibinfo {author} {\bibfnamefont {J.}~\bibnamefont {Skaar}},\ and\ \bibinfo {author} {\bibfnamefont {V.}~\bibnamefont {Makarov}},\ }\bibfield  {title} {\bibinfo {title} {Thermal blinding of gated detectors in quantum cryptography},\ }\href {https://doi.org/10.1364/OE.18.027938} {\bibfield  {journal} {\bibinfo  {journal} {Opt. Express}\ }\textbf {\bibinfo {volume} {18}},\ \bibinfo {pages} {27938} (\bibinfo {year} {2010})}\BibitemShut {NoStop}%
\bibitem [{\citenamefont {Liu}\ \emph {et~al.}(2014)\citenamefont {Liu}, \citenamefont {Lamas-Linares}, \citenamefont {Kurtsiefer}, \citenamefont {Skaar}, \citenamefont {Makarov},\ and\ \citenamefont {Gerhardt}}]{Liu:2014}%
  \BibitemOpen
  \bibfield  {author} {\bibinfo {author} {\bibfnamefont {Q.}~\bibnamefont {Liu}}, \bibinfo {author} {\bibfnamefont {A.}~\bibnamefont {Lamas-Linares}}, \bibinfo {author} {\bibfnamefont {C.}~\bibnamefont {Kurtsiefer}}, \bibinfo {author} {\bibfnamefont {J.}~\bibnamefont {Skaar}}, \bibinfo {author} {\bibfnamefont {V.}~\bibnamefont {Makarov}},\ and\ \bibinfo {author} {\bibfnamefont {I.}~\bibnamefont {Gerhardt}},\ }\bibfield  {title} {\bibinfo {title} {A universal setup for active control of a single-photon detector},\ }\href {https://doi.org/10.1063/1.4854615} {\bibfield  {journal} {\bibinfo  {journal} {Review of Scientific Instruments}\ }\textbf {\bibinfo {volume} {85}},\ \bibinfo {pages} {013108} (\bibinfo {year} {2014})},\ \Eprint {https://arxiv.org/abs/https://pubs.aip.org/aip/rsi/article-pdf/doi/10.1063/1.4854615/14775825/013108\_1\_online.pdf} {https://pubs.aip.org/aip/rsi/article-pdf/doi/10.1063/1.4854615/14775825/013108\_1\_online.pdf} \BibitemShut {NoStop}%
\bibitem [{\citenamefont {Sauge}\ \emph {et~al.}(2011)\citenamefont {Sauge}, \citenamefont {Lydersen}, \citenamefont {Anisimov}, \citenamefont {Skaar},\ and\ \citenamefont {Makarov}}]{Sauge:2011}%
  \BibitemOpen
  \bibfield  {author} {\bibinfo {author} {\bibfnamefont {S.}~\bibnamefont {Sauge}}, \bibinfo {author} {\bibfnamefont {L.}~\bibnamefont {Lydersen}}, \bibinfo {author} {\bibfnamefont {A.}~\bibnamefont {Anisimov}}, \bibinfo {author} {\bibfnamefont {J.}~\bibnamefont {Skaar}},\ and\ \bibinfo {author} {\bibfnamefont {V.}~\bibnamefont {Makarov}},\ }\bibfield  {title} {\bibinfo {title} {Controlling an actively-quenched single photon detector with bright light},\ }\href {https://doi.org/10.1364/OE.19.023590} {\bibfield  {journal} {\bibinfo  {journal} {Opt. Express}\ }\textbf {\bibinfo {volume} {19}},\ \bibinfo {pages} {23590} (\bibinfo {year} {2011})}\BibitemShut {NoStop}%
\bibitem [{\citenamefont {Weier}\ \emph {et~al.}(2011)\citenamefont {Weier}, \citenamefont {Krauss}, \citenamefont {Rau}, \citenamefont {Fürst}, \citenamefont {Nauerth},\ and\ \citenamefont {Weinfurter}}]{Weier:2011}%
  \BibitemOpen
  \bibfield  {author} {\bibinfo {author} {\bibfnamefont {H.}~\bibnamefont {Weier}}, \bibinfo {author} {\bibfnamefont {H.}~\bibnamefont {Krauss}}, \bibinfo {author} {\bibfnamefont {M.}~\bibnamefont {Rau}}, \bibinfo {author} {\bibfnamefont {M.}~\bibnamefont {Fürst}}, \bibinfo {author} {\bibfnamefont {S.}~\bibnamefont {Nauerth}},\ and\ \bibinfo {author} {\bibfnamefont {H.}~\bibnamefont {Weinfurter}},\ }\bibfield  {title} {\bibinfo {title} {Quantum eavesdropping without interception: an attack exploiting the dead time of single-photon detectors},\ }\href {https://doi.org/10.1088/1367-2630/13/7/073024} {\bibfield  {journal} {\bibinfo  {journal} {New Journal of Physics}\ }\textbf {\bibinfo {volume} {13}},\ \bibinfo {pages} {073024} (\bibinfo {year} {2011})}\BibitemShut {NoStop}%
\bibitem [{\citenamefont {Qi}\ \emph {et~al.}(2007)\citenamefont {Qi}, \citenamefont {Fung}, \citenamefont {Lo},\ and\ \citenamefont {Ma}}]{Qi:2007}%
  \BibitemOpen
  \bibfield  {author} {\bibinfo {author} {\bibfnamefont {B.}~\bibnamefont {Qi}}, \bibinfo {author} {\bibfnamefont {C.-H.~F.}\ \bibnamefont {Fung}}, \bibinfo {author} {\bibfnamefont {H.-K.}\ \bibnamefont {Lo}},\ and\ \bibinfo {author} {\bibfnamefont {X.}~\bibnamefont {Ma}},\ }\bibfield  {title} {\bibinfo {title} {Time-shift attack in practical quantum cryptosystems},\ }\href@noop {} {\bibfield  {journal} {\bibinfo  {journal} {Quantum Info. Comput.}\ }\textbf {\bibinfo {volume} {7}},\ \bibinfo {pages} {73–82} (\bibinfo {year} {2007})}\BibitemShut {NoStop}%
\bibitem [{\citenamefont {Yuan}\ \emph {et~al.}(2010)\citenamefont {Yuan}, \citenamefont {Dynes},\ and\ \citenamefont {Shields}}]{Yuan:2010}%
  \BibitemOpen
  \bibfield  {author} {\bibinfo {author} {\bibfnamefont {Z.~L.}\ \bibnamefont {Yuan}}, \bibinfo {author} {\bibfnamefont {J.~F.}\ \bibnamefont {Dynes}},\ and\ \bibinfo {author} {\bibfnamefont {A.~J.}\ \bibnamefont {Shields}},\ }\bibfield  {title} {\bibinfo {title} {Avoiding the blinding attack in qkd},\ }\href {https://doi.org/10.1038/nphoton.2010.269} {\bibfield  {journal} {\bibinfo  {journal} {Nature Photonics}\ }\textbf {\bibinfo {volume} {4}},\ \bibinfo {pages} {800} (\bibinfo {year} {2010})}\BibitemShut {NoStop}%
\bibitem [{\citenamefont {Ferreira~da Silva}\ \emph {et~al.}(2012)\citenamefont {Ferreira~da Silva}, \citenamefont {Xavier}, \citenamefont {Temporão},\ and\ \citenamefont {von~der Weid}}]{FerreiradaSilva:2012}%
  \BibitemOpen
  \bibfield  {author} {\bibinfo {author} {\bibfnamefont {T.}~\bibnamefont {Ferreira~da Silva}}, \bibinfo {author} {\bibfnamefont {G.~B.}\ \bibnamefont {Xavier}}, \bibinfo {author} {\bibfnamefont {G.~P.}\ \bibnamefont {Temporão}},\ and\ \bibinfo {author} {\bibfnamefont {J.~P.}\ \bibnamefont {von~der Weid}},\ }\bibfield  {title} {\bibinfo {title} {Real-time monitoring of single-photon detectors against eavesdropping in quantum key distribution systems},\ }\href {https://doi.org/10.1364/OE.20.018911} {\bibfield  {journal} {\bibinfo  {journal} {Opt. Express}\ }\textbf {\bibinfo {volume} {20}},\ \bibinfo {pages} {18911} (\bibinfo {year} {2012})}\BibitemShut {NoStop}%
\bibitem [{\citenamefont {Ferreira~da Silva}\ \emph {et~al.}(2015)\citenamefont {Ferreira~da Silva}, \citenamefont {do~Amaral}, \citenamefont {Xavier}, \citenamefont {Temporão},\ and\ \citenamefont {von~der Weid}}]{FerreiradaSilva:2014}%
  \BibitemOpen
  \bibfield  {author} {\bibinfo {author} {\bibfnamefont {T.}~\bibnamefont {Ferreira~da Silva}}, \bibinfo {author} {\bibfnamefont {G.~C.}\ \bibnamefont {do~Amaral}}, \bibinfo {author} {\bibfnamefont {G.~B.}\ \bibnamefont {Xavier}}, \bibinfo {author} {\bibfnamefont {G.~P.}\ \bibnamefont {Temporão}},\ and\ \bibinfo {author} {\bibfnamefont {J.~P.}\ \bibnamefont {von~der Weid}},\ }\bibfield  {title} {\bibinfo {title} {Safeguarding quantum key distribution through detection randomization},\ }\href {https://doi.org/10.1109/JSTQE.2014.2361793} {\bibfield  {journal} {\bibinfo  {journal} {IEEE Journal of Selected Topics in Quantum Electronics}\ }\textbf {\bibinfo {volume} {21}},\ \bibinfo {pages} {159} (\bibinfo {year} {2015})}\BibitemShut {NoStop}%
\bibitem [{\citenamefont {Lim}\ \emph {et~al.}(2015)\citenamefont {Lim}, \citenamefont {Walenta}, \citenamefont {Legré}, \citenamefont {Gisin},\ and\ \citenamefont {Zbinden}}]{Lim:2015}%
  \BibitemOpen
  \bibfield  {author} {\bibinfo {author} {\bibfnamefont {C.~C.~W.}\ \bibnamefont {Lim}}, \bibinfo {author} {\bibfnamefont {N.}~\bibnamefont {Walenta}}, \bibinfo {author} {\bibfnamefont {M.}~\bibnamefont {Legré}}, \bibinfo {author} {\bibfnamefont {N.}~\bibnamefont {Gisin}},\ and\ \bibinfo {author} {\bibfnamefont {H.}~\bibnamefont {Zbinden}},\ }\bibfield  {title} {\bibinfo {title} {Random variation of detector efficiency: A countermeasure against detector blinding attacks for quantum key distribution},\ }\href {https://doi.org/10.1109/JSTQE.2015.2389528} {\bibfield  {journal} {\bibinfo  {journal} {IEEE Journal of Selected Topics in Quantum Electronics}\ }\textbf {\bibinfo {volume} {21}},\ \bibinfo {pages} {192} (\bibinfo {year} {2015})}\BibitemShut {NoStop}%
\bibitem [{\citenamefont {Qian}\ \emph {et~al.}(2019)\citenamefont {Qian}, \citenamefont {He}, \citenamefont {Wang}, \citenamefont {Chen}, \citenamefont {Yin}, \citenamefont {Guo},\ and\ \citenamefont {Han}}]{Qian:2019}%
  \BibitemOpen
  \bibfield  {author} {\bibinfo {author} {\bibfnamefont {Y.-J.}\ \bibnamefont {Qian}}, \bibinfo {author} {\bibfnamefont {D.-Y.}\ \bibnamefont {He}}, \bibinfo {author} {\bibfnamefont {S.}~\bibnamefont {Wang}}, \bibinfo {author} {\bibfnamefont {W.}~\bibnamefont {Chen}}, \bibinfo {author} {\bibfnamefont {Z.-Q.}\ \bibnamefont {Yin}}, \bibinfo {author} {\bibfnamefont {G.-C.}\ \bibnamefont {Guo}},\ and\ \bibinfo {author} {\bibfnamefont {Z.-F.}\ \bibnamefont {Han}},\ }\bibfield  {title} {\bibinfo {title} {Robust countermeasure against detector control attack in a practical quantum key distribution system},\ }\href {https://doi.org/10.1364/OPTICA.6.001178} {\bibfield  {journal} {\bibinfo  {journal} {Optica}\ }\textbf {\bibinfo {volume} {6}},\ \bibinfo {pages} {1178} (\bibinfo {year} {2019})}\BibitemShut {NoStop}%
\bibitem [{\citenamefont {{Acheva, Polina}}\ \emph {et~al.}(2023)\citenamefont {{Acheva, Polina}}, \citenamefont {{Zaitsev, Konstantin}}, \citenamefont {{Zavodilenko, Vladimir}}, \citenamefont {{Losev, Anton}}, \citenamefont {{Huang, Anqi}},\ and\ \citenamefont {{Makarov, Vadim}}}]{Acheva:2023}%
  \BibitemOpen
  \bibfield  {author} {\bibinfo {author} {\bibnamefont {{Acheva, Polina}}}, \bibinfo {author} {\bibnamefont {{Zaitsev, Konstantin}}}, \bibinfo {author} {\bibnamefont {{Zavodilenko, Vladimir}}}, \bibinfo {author} {\bibnamefont {{Losev, Anton}}}, \bibinfo {author} {\bibnamefont {{Huang, Anqi}}},\ and\ \bibinfo {author} {\bibnamefont {{Makarov, Vadim}}},\ }\bibfield  {title} {\bibinfo {title} {Automated verification of countermeasure against detector-control attack in quantum key distribution},\ }\href {https://doi.org/10.1140/epjqt/s40507-023-00178-x} {\bibfield  {journal} {\bibinfo  {journal} {EPJ Quantum Technol.}\ }\textbf {\bibinfo {volume} {10}},\ \bibinfo {pages} {22} (\bibinfo {year} {2023})}\BibitemShut {NoStop}%
\bibitem [{\citenamefont {Lo}\ \emph {et~al.}(2012)\citenamefont {Lo}, \citenamefont {Curty},\ and\ \citenamefont {Qi}}]{Lo:2012}%
  \BibitemOpen
  \bibfield  {author} {\bibinfo {author} {\bibfnamefont {H.-K.}\ \bibnamefont {Lo}}, \bibinfo {author} {\bibfnamefont {M.}~\bibnamefont {Curty}},\ and\ \bibinfo {author} {\bibfnamefont {B.}~\bibnamefont {Qi}},\ }\bibfield  {title} {\bibinfo {title} {Measurement-device-independent quantum key distribution},\ }\href@noop {} {\bibfield  {journal} {\bibinfo  {journal} {Phys. Rev. Lett.}\ }\textbf {\bibinfo {volume} {108}},\ \bibinfo {pages} {130503} (\bibinfo {year} {2012})}\BibitemShut {NoStop}%
\bibitem [{\citenamefont {Lucamarini}\ \emph {et~al.}(2018)\citenamefont {Lucamarini}, \citenamefont {Yuan}, \citenamefont {Dynes},\ and\ \citenamefont {Shields}}]{Lucamarini:2018}%
  \BibitemOpen
  \bibfield  {author} {\bibinfo {author} {\bibfnamefont {M.}~\bibnamefont {Lucamarini}}, \bibinfo {author} {\bibfnamefont {Z.~L.}\ \bibnamefont {Yuan}}, \bibinfo {author} {\bibfnamefont {J.~F.}\ \bibnamefont {Dynes}},\ and\ \bibinfo {author} {\bibfnamefont {A.~J.}\ \bibnamefont {Shields}},\ }\bibfield  {title} {\bibinfo {title} {Overcoming the rate--distance limit of quantum key distribution without quantum repeaters},\ }\href {https://doi.org/10.1038/s41586-018-0066-6} {\bibfield  {journal} {\bibinfo  {journal} {Nature}\ }\textbf {\bibinfo {volume} {557}},\ \bibinfo {pages} {400} (\bibinfo {year} {2018})}\BibitemShut {NoStop}%
\bibitem [{\citenamefont {Yin}\ and\ \citenamefont {Fu}(2019)}]{Yin:2019}%
  \BibitemOpen
  \bibfield  {author} {\bibinfo {author} {\bibfnamefont {H.-L.}\ \bibnamefont {Yin}}\ and\ \bibinfo {author} {\bibfnamefont {Y.}~\bibnamefont {Fu}},\ }\bibfield  {title} {\bibinfo {title} {Measurement-device-independent twin-field quantum key distribution},\ }\href {https://doi.org/10.1038/s41598-019-39454-1} {\bibfield  {journal} {\bibinfo  {journal} {Scientific Reports}\ }\textbf {\bibinfo {volume} {9}},\ \bibinfo {pages} {3045} (\bibinfo {year} {2019})}\BibitemShut {NoStop}%
\bibitem [{\citenamefont {Pirandola}\ \emph {et~al.}(2017)\citenamefont {Pirandola}, \citenamefont {Laurenza}, \citenamefont {Ottaviani},\ and\ \citenamefont {Banchi}}]{Pirandola:2017}%
  \BibitemOpen
  \bibfield  {author} {\bibinfo {author} {\bibfnamefont {S.}~\bibnamefont {Pirandola}}, \bibinfo {author} {\bibfnamefont {R.}~\bibnamefont {Laurenza}}, \bibinfo {author} {\bibfnamefont {C.}~\bibnamefont {Ottaviani}},\ and\ \bibinfo {author} {\bibfnamefont {L.}~\bibnamefont {Banchi}},\ }\bibfield  {title} {\bibinfo {title} {Fundamental limits of repeaterless quantum communications},\ }\href {https://doi.org/10.1038/ncomms15043} {\bibfield  {journal} {\bibinfo  {journal} {Nature Communications}\ }\textbf {\bibinfo {volume} {8}},\ \bibinfo {pages} {15043} (\bibinfo {year} {2017})}\BibitemShut {NoStop}%
\bibitem [{\citenamefont {Chen}\ \emph {et~al.}(2021)\citenamefont {Chen}, \citenamefont {Zhang}, \citenamefont {Liu}, \citenamefont {Jiang}, \citenamefont {Zhang}, \citenamefont {Han}, \citenamefont {Ma}, \citenamefont {Hu}, \citenamefont {Li}, \citenamefont {Liu}, \citenamefont {Zhou}, \citenamefont {Jiang}, \citenamefont {Chen}, \citenamefont {Li}, \citenamefont {You}, \citenamefont {Wang}, \citenamefont {Wang}, \citenamefont {Zhang},\ and\ \citenamefont {Pan}}]{Chen:2021}%
  \BibitemOpen
  \bibfield  {author} {\bibinfo {author} {\bibfnamefont {J.-P.}\ \bibnamefont {Chen}}, \bibinfo {author} {\bibfnamefont {C.}~\bibnamefont {Zhang}}, \bibinfo {author} {\bibfnamefont {Y.}~\bibnamefont {Liu}}, \bibinfo {author} {\bibfnamefont {C.}~\bibnamefont {Jiang}}, \bibinfo {author} {\bibfnamefont {W.-J.}\ \bibnamefont {Zhang}}, \bibinfo {author} {\bibfnamefont {Z.-Y.}\ \bibnamefont {Han}}, \bibinfo {author} {\bibfnamefont {S.-Z.}\ \bibnamefont {Ma}}, \bibinfo {author} {\bibfnamefont {X.-L.}\ \bibnamefont {Hu}}, \bibinfo {author} {\bibfnamefont {Y.-H.}\ \bibnamefont {Li}}, \bibinfo {author} {\bibfnamefont {H.}~\bibnamefont {Liu}}, \bibinfo {author} {\bibfnamefont {F.}~\bibnamefont {Zhou}}, \bibinfo {author} {\bibfnamefont {H.-F.}\ \bibnamefont {Jiang}}, \bibinfo {author} {\bibfnamefont {T.-Y.}\ \bibnamefont {Chen}}, \bibinfo {author} {\bibfnamefont {H.}~\bibnamefont {Li}}, \bibinfo {author} {\bibfnamefont {L.-X.}\ \bibnamefont {You}}, \bibinfo {author} {\bibfnamefont {Z.}~\bibnamefont {Wang}}, \bibinfo
  {author} {\bibfnamefont {X.-B.}\ \bibnamefont {Wang}}, \bibinfo {author} {\bibfnamefont {Q.}~\bibnamefont {Zhang}},\ and\ \bibinfo {author} {\bibfnamefont {J.-W.}\ \bibnamefont {Pan}},\ }\bibfield  {title} {\bibinfo {title} {Twin-field quantum key distribution over a 511 km optical fibre linking two distant metropolitan areas},\ }\href {https://doi.org/10.1038/s41566-021-00828-5} {\bibfield  {journal} {\bibinfo  {journal} {Nature Photonics}\ }\textbf {\bibinfo {volume} {15}},\ \bibinfo {pages} {570} (\bibinfo {year} {2021})}\BibitemShut {NoStop}%
\bibitem [{\citenamefont {Liu}\ \emph {et~al.}(2021)\citenamefont {Liu}, \citenamefont {Jiang}, \citenamefont {Zhu}, \citenamefont {Zou}, \citenamefont {Yu}, \citenamefont {Hu}, \citenamefont {Xu}, \citenamefont {Ma}, \citenamefont {Han}, \citenamefont {Chen}, \citenamefont {Dai}, \citenamefont {Tang}, \citenamefont {Zhang}, \citenamefont {Li}, \citenamefont {You}, \citenamefont {Wang}, \citenamefont {Hua}, \citenamefont {Hu}, \citenamefont {Zhang}, \citenamefont {Zhou}, \citenamefont {Zhang}, \citenamefont {Wang}, \citenamefont {Chen},\ and\ \citenamefont {Pan}}]{Liu:2021}%
  \BibitemOpen
  \bibfield  {author} {\bibinfo {author} {\bibfnamefont {H.}~\bibnamefont {Liu}}, \bibinfo {author} {\bibfnamefont {C.}~\bibnamefont {Jiang}}, \bibinfo {author} {\bibfnamefont {H.-T.}\ \bibnamefont {Zhu}}, \bibinfo {author} {\bibfnamefont {M.}~\bibnamefont {Zou}}, \bibinfo {author} {\bibfnamefont {Z.-W.}\ \bibnamefont {Yu}}, \bibinfo {author} {\bibfnamefont {X.-L.}\ \bibnamefont {Hu}}, \bibinfo {author} {\bibfnamefont {H.}~\bibnamefont {Xu}}, \bibinfo {author} {\bibfnamefont {S.}~\bibnamefont {Ma}}, \bibinfo {author} {\bibfnamefont {Z.}~\bibnamefont {Han}}, \bibinfo {author} {\bibfnamefont {J.-P.}\ \bibnamefont {Chen}}, \bibinfo {author} {\bibfnamefont {Y.}~\bibnamefont {Dai}}, \bibinfo {author} {\bibfnamefont {S.-B.}\ \bibnamefont {Tang}}, \bibinfo {author} {\bibfnamefont {W.}~\bibnamefont {Zhang}}, \bibinfo {author} {\bibfnamefont {H.}~\bibnamefont {Li}}, \bibinfo {author} {\bibfnamefont {L.}~\bibnamefont {You}}, \bibinfo {author} {\bibfnamefont {Z.}~\bibnamefont {Wang}}, \bibinfo {author} {\bibfnamefont
  {Y.}~\bibnamefont {Hua}}, \bibinfo {author} {\bibfnamefont {H.}~\bibnamefont {Hu}}, \bibinfo {author} {\bibfnamefont {H.}~\bibnamefont {Zhang}}, \bibinfo {author} {\bibfnamefont {F.}~\bibnamefont {Zhou}}, \bibinfo {author} {\bibfnamefont {Q.}~\bibnamefont {Zhang}}, \bibinfo {author} {\bibfnamefont {X.-B.}\ \bibnamefont {Wang}}, \bibinfo {author} {\bibfnamefont {T.-Y.}\ \bibnamefont {Chen}},\ and\ \bibinfo {author} {\bibfnamefont {J.-W.}\ \bibnamefont {Pan}},\ }\bibfield  {title} {\bibinfo {title} {Field test of twin-field quantum key distribution through sending-or-not-sending over 428 km},\ }\href {https://doi.org/10.1103/PhysRevLett.126.250502} {\bibfield  {journal} {\bibinfo  {journal} {Phys. Rev. Lett.}\ }\textbf {\bibinfo {volume} {126}},\ \bibinfo {pages} {250502} (\bibinfo {year} {2021})}\BibitemShut {NoStop}%
\bibitem [{\citenamefont {Wang}\ \emph {et~al.}(2022)\citenamefont {Wang}, \citenamefont {Yin}, \citenamefont {He}, \citenamefont {Chen}, \citenamefont {Wang}, \citenamefont {Ye}, \citenamefont {Zhou}, \citenamefont {Fan-Yuan}, \citenamefont {Wang}, \citenamefont {Zhu}, \citenamefont {Morozov}, \citenamefont {Divochiy}, \citenamefont {Zhou}, \citenamefont {Guo},\ and\ \citenamefont {Han}}]{Wang:2022}%
  \BibitemOpen
  \bibfield  {author} {\bibinfo {author} {\bibfnamefont {S.}~\bibnamefont {Wang}}, \bibinfo {author} {\bibfnamefont {Z.-Q.}\ \bibnamefont {Yin}}, \bibinfo {author} {\bibfnamefont {D.-Y.}\ \bibnamefont {He}}, \bibinfo {author} {\bibfnamefont {W.}~\bibnamefont {Chen}}, \bibinfo {author} {\bibfnamefont {R.-Q.}\ \bibnamefont {Wang}}, \bibinfo {author} {\bibfnamefont {P.}~\bibnamefont {Ye}}, \bibinfo {author} {\bibfnamefont {Y.}~\bibnamefont {Zhou}}, \bibinfo {author} {\bibfnamefont {G.-J.}\ \bibnamefont {Fan-Yuan}}, \bibinfo {author} {\bibfnamefont {F.-X.}\ \bibnamefont {Wang}}, \bibinfo {author} {\bibfnamefont {Y.-G.}\ \bibnamefont {Zhu}}, \bibinfo {author} {\bibfnamefont {P.~V.}\ \bibnamefont {Morozov}}, \bibinfo {author} {\bibfnamefont {A.~V.}\ \bibnamefont {Divochiy}}, \bibinfo {author} {\bibfnamefont {Z.}~\bibnamefont {Zhou}}, \bibinfo {author} {\bibfnamefont {G.-C.}\ \bibnamefont {Guo}},\ and\ \bibinfo {author} {\bibfnamefont {Z.-F.}\ \bibnamefont {Han}},\ }\bibfield  {title} {\bibinfo {title} {Twin-field
  quantum key distribution over 830-km fibre},\ }\href {https://doi.org/10.1038/s41566-021-00928-2} {\bibfield  {journal} {\bibinfo  {journal} {Nature Photonics}\ }\textbf {\bibinfo {volume} {16}},\ \bibinfo {pages} {154} (\bibinfo {year} {2022})}\BibitemShut {NoStop}%
\bibitem [{\citenamefont {Liu}\ \emph {et~al.}(2023)\citenamefont {Liu}, \citenamefont {Zhang}, \citenamefont {Jiang}, \citenamefont {Chen}, \citenamefont {Zhang}, \citenamefont {Pan}, \citenamefont {Ma}, \citenamefont {Dong}, \citenamefont {Xiong}, \citenamefont {Zhang}, \citenamefont {Li}, \citenamefont {Wang}, \citenamefont {Wu}, \citenamefont {Chen}, \citenamefont {You}, \citenamefont {Wang}, \citenamefont {Zhang},\ and\ \citenamefont {Pan}}]{Yang:2023}%
  \BibitemOpen
  \bibfield  {author} {\bibinfo {author} {\bibfnamefont {Y.}~\bibnamefont {Liu}}, \bibinfo {author} {\bibfnamefont {W.-J.}\ \bibnamefont {Zhang}}, \bibinfo {author} {\bibfnamefont {C.}~\bibnamefont {Jiang}}, \bibinfo {author} {\bibfnamefont {J.-P.}\ \bibnamefont {Chen}}, \bibinfo {author} {\bibfnamefont {C.}~\bibnamefont {Zhang}}, \bibinfo {author} {\bibfnamefont {W.-X.}\ \bibnamefont {Pan}}, \bibinfo {author} {\bibfnamefont {D.}~\bibnamefont {Ma}}, \bibinfo {author} {\bibfnamefont {H.}~\bibnamefont {Dong}}, \bibinfo {author} {\bibfnamefont {J.-M.}\ \bibnamefont {Xiong}}, \bibinfo {author} {\bibfnamefont {C.-J.}\ \bibnamefont {Zhang}}, \bibinfo {author} {\bibfnamefont {H.}~\bibnamefont {Li}}, \bibinfo {author} {\bibfnamefont {R.-C.}\ \bibnamefont {Wang}}, \bibinfo {author} {\bibfnamefont {J.}~\bibnamefont {Wu}}, \bibinfo {author} {\bibfnamefont {T.-Y.}\ \bibnamefont {Chen}}, \bibinfo {author} {\bibfnamefont {L.}~\bibnamefont {You}}, \bibinfo {author} {\bibfnamefont {X.-B.}\ \bibnamefont {Wang}}, \bibinfo
  {author} {\bibfnamefont {Q.}~\bibnamefont {Zhang}},\ and\ \bibinfo {author} {\bibfnamefont {J.-W.}\ \bibnamefont {Pan}},\ }\bibfield  {title} {\bibinfo {title} {Experimental twin-field quantum key distribution over 1000 km fiber distance},\ }\href {https://doi.org/10.1103/PhysRevLett.130.210801} {\bibfield  {journal} {\bibinfo  {journal} {Phys. Rev. Lett.}\ }\textbf {\bibinfo {volume} {130}},\ \bibinfo {pages} {210801} (\bibinfo {year} {2023})}\BibitemShut {NoStop}%
\bibitem [{\citenamefont {Zhou}\ \emph {et~al.}(2023{\natexlab{a}})\citenamefont {Zhou}, \citenamefont {Lin}, \citenamefont {Jing},\ and\ \citenamefont {Yuan}}]{Yuan:2022}%
  \BibitemOpen
  \bibfield  {author} {\bibinfo {author} {\bibfnamefont {L.}~\bibnamefont {Zhou}}, \bibinfo {author} {\bibfnamefont {J.}~\bibnamefont {Lin}}, \bibinfo {author} {\bibfnamefont {Y.}~\bibnamefont {Jing}},\ and\ \bibinfo {author} {\bibfnamefont {Z.}~\bibnamefont {Yuan}},\ }\bibfield  {title} {\bibinfo {title} {Twin-field quantum key distribution without optical frequency dissemination},\ }\href {https://doi.org/10.1038/s41467-023-36573-2} {\bibfield  {journal} {\bibinfo  {journal} {Nature Communications}\ }\textbf {\bibinfo {volume} {14}},\ \bibinfo {pages} {928} (\bibinfo {year} {2023}{\natexlab{a}})}\BibitemShut {NoStop}%
\bibitem [{\citenamefont {Li}\ \emph {et~al.}(2023)\citenamefont {Li}, \citenamefont {Zhang}, \citenamefont {Lu}, \citenamefont {Li}, \citenamefont {Jiang}, \citenamefont {Liu}, \citenamefont {Huang}, \citenamefont {Li}, \citenamefont {Wang}, \citenamefont {Wang}, \citenamefont {Zhang}, \citenamefont {You}, \citenamefont {Xu},\ and\ \citenamefont {Pan}}]{Wei:2023}%
  \BibitemOpen
  \bibfield  {author} {\bibinfo {author} {\bibfnamefont {W.}~\bibnamefont {Li}}, \bibinfo {author} {\bibfnamefont {L.}~\bibnamefont {Zhang}}, \bibinfo {author} {\bibfnamefont {Y.}~\bibnamefont {Lu}}, \bibinfo {author} {\bibfnamefont {Z.-P.}\ \bibnamefont {Li}}, \bibinfo {author} {\bibfnamefont {C.}~\bibnamefont {Jiang}}, \bibinfo {author} {\bibfnamefont {Y.}~\bibnamefont {Liu}}, \bibinfo {author} {\bibfnamefont {J.}~\bibnamefont {Huang}}, \bibinfo {author} {\bibfnamefont {H.}~\bibnamefont {Li}}, \bibinfo {author} {\bibfnamefont {Z.}~\bibnamefont {Wang}}, \bibinfo {author} {\bibfnamefont {X.-B.}\ \bibnamefont {Wang}}, \bibinfo {author} {\bibfnamefont {Q.}~\bibnamefont {Zhang}}, \bibinfo {author} {\bibfnamefont {L.}~\bibnamefont {You}}, \bibinfo {author} {\bibfnamefont {F.}~\bibnamefont {Xu}},\ and\ \bibinfo {author} {\bibfnamefont {J.-W.}\ \bibnamefont {Pan}},\ }\bibfield  {title} {\bibinfo {title} {Twin-field quantum key distribution without phase locking},\ }\href
  {https://doi.org/10.1103/PhysRevLett.130.250802} {\bibfield  {journal} {\bibinfo  {journal} {Phys. Rev. Lett.}\ }\textbf {\bibinfo {volume} {130}},\ \bibinfo {pages} {250802} (\bibinfo {year} {2023})}\BibitemShut {NoStop}%
\bibitem [{\citenamefont {Zhou}\ \emph {et~al.}(2023{\natexlab{b}})\citenamefont {Zhou}, \citenamefont {Lin}, \citenamefont {Xie}, \citenamefont {Lu}, \citenamefont {Jing}, \citenamefont {Yin},\ and\ \citenamefont {Yuan}}]{Zhou:2023}%
  \BibitemOpen
  \bibfield  {author} {\bibinfo {author} {\bibfnamefont {L.}~\bibnamefont {Zhou}}, \bibinfo {author} {\bibfnamefont {J.}~\bibnamefont {Lin}}, \bibinfo {author} {\bibfnamefont {Y.-M.}\ \bibnamefont {Xie}}, \bibinfo {author} {\bibfnamefont {Y.-S.}\ \bibnamefont {Lu}}, \bibinfo {author} {\bibfnamefont {Y.}~\bibnamefont {Jing}}, \bibinfo {author} {\bibfnamefont {H.-L.}\ \bibnamefont {Yin}},\ and\ \bibinfo {author} {\bibfnamefont {Z.}~\bibnamefont {Yuan}},\ }\bibfield  {title} {\bibinfo {title} {Experimental quantum communication overcomes the rate-loss limit without global phase tracking},\ }\href {https://doi.org/10.1103/PhysRevLett.130.250801} {\bibfield  {journal} {\bibinfo  {journal} {Phys. Rev. Lett.}\ }\textbf {\bibinfo {volume} {130}},\ \bibinfo {pages} {250801} (\bibinfo {year} {2023}{\natexlab{b}})}\BibitemShut {NoStop}%
\bibitem [{\citenamefont {Zhong}\ \emph {et~al.}(2019)\citenamefont {Zhong}, \citenamefont {Hu}, \citenamefont {Curty}, \citenamefont {Qian},\ and\ \citenamefont {Lo}}]{Zhong:2019}%
  \BibitemOpen
  \bibfield  {author} {\bibinfo {author} {\bibfnamefont {X.}~\bibnamefont {Zhong}}, \bibinfo {author} {\bibfnamefont {J.}~\bibnamefont {Hu}}, \bibinfo {author} {\bibfnamefont {M.}~\bibnamefont {Curty}}, \bibinfo {author} {\bibfnamefont {L.}~\bibnamefont {Qian}},\ and\ \bibinfo {author} {\bibfnamefont {H.-K.}\ \bibnamefont {Lo}},\ }\bibfield  {title} {\bibinfo {title} {Proof-of-principle experimental demonstration of twin-field type quantum key distribution},\ }\href {https://doi.org/10.1103/PhysRevLett.123.100506} {\bibfield  {journal} {\bibinfo  {journal} {Phys. Rev. Lett.}\ }\textbf {\bibinfo {volume} {123}},\ \bibinfo {pages} {100506} (\bibinfo {year} {2019})}\BibitemShut {NoStop}%
\bibitem [{\citenamefont {Zhong}\ \emph {et~al.}(2022)\citenamefont {Zhong}, \citenamefont {Wang}, \citenamefont {Mandil}, \citenamefont {Lo},\ and\ \citenamefont {Qian}}]{Zhong:2022}%
  \BibitemOpen
  \bibfield  {author} {\bibinfo {author} {\bibfnamefont {X.}~\bibnamefont {Zhong}}, \bibinfo {author} {\bibfnamefont {W.}~\bibnamefont {Wang}}, \bibinfo {author} {\bibfnamefont {R.}~\bibnamefont {Mandil}}, \bibinfo {author} {\bibfnamefont {H.-K.}\ \bibnamefont {Lo}},\ and\ \bibinfo {author} {\bibfnamefont {L.}~\bibnamefont {Qian}},\ }\bibfield  {title} {\bibinfo {title} {Simple multiuser twin-field quantum key distribution network},\ }\href {https://doi.org/10.1103/PhysRevApplied.17.014025} {\bibfield  {journal} {\bibinfo  {journal} {Phys. Rev. Appl.}\ }\textbf {\bibinfo {volume} {17}},\ \bibinfo {pages} {014025} (\bibinfo {year} {2022})}\BibitemShut {NoStop}%
\bibitem [{\citenamefont {Mandil}\ \emph {et~al.}(2025)\citenamefont {Mandil}, \citenamefont {Qian},\ and\ \citenamefont {Lo}}]{Mandil:2025}%
  \BibitemOpen
  \bibfield  {author} {\bibinfo {author} {\bibfnamefont {R.}~\bibnamefont {Mandil}}, \bibinfo {author} {\bibfnamefont {L.}~\bibnamefont {Qian}},\ and\ \bibinfo {author} {\bibfnamefont {H.-K.}\ \bibnamefont {Lo}},\ }\bibfield  {title} {\bibinfo {title} {Long-fiber sagnac interferometers for twin-field quantum key distribution networks},\ }\href {https://doi.org/10.1103/PhysRevApplied.23.034040} {\bibfield  {journal} {\bibinfo  {journal} {Phys. Rev. Appl.}\ }\textbf {\bibinfo {volume} {23}},\ \bibinfo {pages} {034040} (\bibinfo {year} {2025})}\BibitemShut {NoStop}%
\bibitem [{\citenamefont {Xavier}\ \emph {et~al.}(2011{\natexlab{a}})\citenamefont {Xavier}, \citenamefont {da~Silva}, \citenamefont {Temporão},\ and\ \citenamefont {von~der Weid}}]{Xavier:2011a}%
  \BibitemOpen
  \bibfield  {author} {\bibinfo {author} {\bibfnamefont {G.}~\bibnamefont {Xavier}}, \bibinfo {author} {\bibfnamefont {T.}~\bibnamefont {da~Silva}}, \bibinfo {author} {\bibfnamefont {G.}~\bibnamefont {Temporão}},\ and\ \bibinfo {author} {\bibfnamefont {J.}~\bibnamefont {von~der Weid}},\ }\bibfield  {title} {\bibinfo {title} {Polarisation drift compensation in 8km-long mach-zehnder fibre-optical interferometer for quantum communication},\ }\href {https://doi.org/10.1049/el.2011.0470} {\bibfield  {journal} {\bibinfo  {journal} {Electronics Letters}\ }\textbf {\bibinfo {volume} {47}},\ \bibinfo {pages} {608} (\bibinfo {year} {2011}{\natexlab{a}})},\ \Eprint {https://arxiv.org/abs/https://digital-library.theiet.org/doi/pdf/10.1049/el.2011.0470} {https://digital-library.theiet.org/doi/pdf/10.1049/el.2011.0470} \BibitemShut {NoStop}%
\bibitem [{\citenamefont {Xavier}\ \emph {et~al.}(2011{\natexlab{b}})\citenamefont {Xavier}, \citenamefont {de~Faria}, \citenamefont {da~Silva}, \citenamefont {Temporão},\ and\ \citenamefont {von~der Weid}}]{Xavier:2011b}%
  \BibitemOpen
  \bibfield  {author} {\bibinfo {author} {\bibfnamefont {G.~B.}\ \bibnamefont {Xavier}}, \bibinfo {author} {\bibfnamefont {G.~V.}\ \bibnamefont {de~Faria}}, \bibinfo {author} {\bibfnamefont {T.~F.}\ \bibnamefont {da~Silva}}, \bibinfo {author} {\bibfnamefont {G.~P.}\ \bibnamefont {Temporão}},\ and\ \bibinfo {author} {\bibfnamefont {J.~P.}\ \bibnamefont {von~der Weid}},\ }\bibfield  {title} {\bibinfo {title} {Active polarization control for quantum communication in long-distance optical fibers with shared telecom traffic},\ }\href {https://doi.org/https://doi.org/10.1002/mop.26320} {\bibfield  {journal} {\bibinfo  {journal} {Microwave and Optical Technology Letters}\ }\textbf {\bibinfo {volume} {53}},\ \bibinfo {pages} {2661} (\bibinfo {year} {2011}{\natexlab{b}})}\BibitemShut {NoStop}%
\bibitem [{\citenamefont {Xavier}\ \emph {et~al.}(2008)\citenamefont {Xavier}, \citenamefont {de~Faria}, \citenamefont {Temporão},\ and\ \citenamefont {von~der Weid}}]{Xavier:08}%
  \BibitemOpen
  \bibfield  {author} {\bibinfo {author} {\bibfnamefont {G.~B.}\ \bibnamefont {Xavier}}, \bibinfo {author} {\bibfnamefont {G.~V.}\ \bibnamefont {de~Faria}}, \bibinfo {author} {\bibfnamefont {G.~P.}\ \bibnamefont {Temporão}},\ and\ \bibinfo {author} {\bibfnamefont {J.~P.}\ \bibnamefont {von~der Weid}},\ }\bibfield  {title} {\bibinfo {title} {Full polarization control for fiber optical quantum communication systems using polarization encoding},\ }\href {https://doi.org/10.1364/OE.16.001867} {\bibfield  {journal} {\bibinfo  {journal} {Opt. Express}\ }\textbf {\bibinfo {volume} {16}},\ \bibinfo {pages} {1867} (\bibinfo {year} {2008})}\BibitemShut {NoStop}%
\bibitem [{\citenamefont {de~Faria}\ \emph {et~al.}(2008)\citenamefont {de~Faria}, \citenamefont {Ferreira}, \citenamefont {Xavier}, \citenamefont {Temporão},\ and\ \citenamefont {von~der Weid}}]{Faria:2008}%
  \BibitemOpen
  \bibfield  {author} {\bibinfo {author} {\bibfnamefont {G.~V.}\ \bibnamefont {de~Faria}}, \bibinfo {author} {\bibfnamefont {J.}~\bibnamefont {Ferreira}}, \bibinfo {author} {\bibfnamefont {G.}~\bibnamefont {Xavier}}, \bibinfo {author} {\bibfnamefont {G.}~\bibnamefont {Temporão}},\ and\ \bibinfo {author} {\bibfnamefont {J.}~\bibnamefont {von~der Weid}},\ }\bibfield  {title} {\bibinfo {title} {Polarisation control schemes for fibre-optics quantum communications using polarisation encoding},\ }\href {https://doi.org/10.1049/el:20083122} {\bibfield  {journal} {\bibinfo  {journal} {Electronics Letters}\ }\textbf {\bibinfo {volume} {44}},\ \bibinfo {pages} {228} (\bibinfo {year} {2008})}\BibitemShut {NoStop}%
\bibitem [{\citenamefont {Ramos}\ \emph {et~al.}(2022)\citenamefont {Ramos}, \citenamefont {Silva}, \citenamefont {Muga},\ and\ \citenamefont {Pinto}}]{Ramos:2022}%
  \BibitemOpen
  \bibfield  {author} {\bibinfo {author} {\bibfnamefont {M.~F.}\ \bibnamefont {Ramos}}, \bibinfo {author} {\bibfnamefont {N.~A.}\ \bibnamefont {Silva}}, \bibinfo {author} {\bibfnamefont {N.~J.}\ \bibnamefont {Muga}},\ and\ \bibinfo {author} {\bibfnamefont {A.~N.}\ \bibnamefont {Pinto}},\ }\bibfield  {title} {\bibinfo {title} {Full polarization random drift compensation method for quantum communication},\ }\href {https://doi.org/10.1364/OE.445228} {\bibfield  {journal} {\bibinfo  {journal} {Opt. Express}\ }\textbf {\bibinfo {volume} {30}},\ \bibinfo {pages} {6907} (\bibinfo {year} {2022})}\BibitemShut {NoStop}%
\bibitem [{\citenamefont {Ribordy}\ \emph {et~al.}(1998)\citenamefont {Ribordy}, \citenamefont {Gautier}, \citenamefont {Gisin}, \citenamefont {Guinnard},\ and\ \citenamefont {Zbinden}}]{Ribordy:1998}%
  \BibitemOpen
  \bibfield  {author} {\bibinfo {author} {\bibfnamefont {G.}~\bibnamefont {Ribordy}}, \bibinfo {author} {\bibfnamefont {J.-D.}\ \bibnamefont {Gautier}}, \bibinfo {author} {\bibfnamefont {N.}~\bibnamefont {Gisin}}, \bibinfo {author} {\bibfnamefont {O.}~\bibnamefont {Guinnard}},\ and\ \bibinfo {author} {\bibfnamefont {H.}~\bibnamefont {Zbinden}},\ }\bibfield  {title} {\bibinfo {title} {Automated plug and play quantum key distribution},\ }\href {https://doi.org/10.1049/el:19981473} {\bibfield  {journal} {\bibinfo  {journal} {Electronics Letters}\ }\textbf {\bibinfo {volume} {34}},\ \bibinfo {pages} {2116} (\bibinfo {year} {1998})}\BibitemShut {NoStop}%
\bibitem [{\citenamefont {Xue}\ \emph {et~al.}(2021)\citenamefont {Xue}, \citenamefont {Zhao}, \citenamefont {Mao},\ and\ \citenamefont {Xu}}]{Xue:2021}%
  \BibitemOpen
  \bibfield  {author} {\bibinfo {author} {\bibfnamefont {K.}~\bibnamefont {Xue}}, \bibinfo {author} {\bibfnamefont {S.}~\bibnamefont {Zhao}}, \bibinfo {author} {\bibfnamefont {Q.}~\bibnamefont {Mao}},\ and\ \bibinfo {author} {\bibfnamefont {R.}~\bibnamefont {Xu}},\ }\bibfield  {title} {\bibinfo {title} {Plug-and-play sending-or-not-sending twin-field quantum key distribution},\ }\href {https://doi.org/10.1007/s11128-021-03259-x} {\bibfield  {journal} {\bibinfo  {journal} {Quantum Information Processing}\ }\textbf {\bibinfo {volume} {20}},\ \bibinfo {pages} {320} (\bibinfo {year} {2021})}\BibitemShut {NoStop}%
\bibitem [{\citenamefont {Calliari}\ \emph {et~al.}(2019)\citenamefont {Calliari}, \citenamefont {Tovar}, \citenamefont {Nascimento}, \citenamefont {Perlingeiro}, \citenamefont {Amaral},\ and\ \citenamefont {Temporão}}]{Calliari:2019}%
  \BibitemOpen
  \bibfield  {author} {\bibinfo {author} {\bibfnamefont {F.}~\bibnamefont {Calliari}}, \bibinfo {author} {\bibfnamefont {P.}~\bibnamefont {Tovar}}, \bibinfo {author} {\bibfnamefont {C.}~\bibnamefont {Nascimento}}, \bibinfo {author} {\bibfnamefont {B.}~\bibnamefont {Perlingeiro}}, \bibinfo {author} {\bibfnamefont {G.}~\bibnamefont {Amaral}},\ and\ \bibinfo {author} {\bibfnamefont {G.}~\bibnamefont {Temporão}},\ }\bibfield  {title} {\bibinfo {title} {Alignment-free characterization of polarizing beamsplitters},\ }\href {https://doi.org/10.1364/AO.58.004395} {\bibfield  {journal} {\bibinfo  {journal} {Appl. Opt.}\ }\textbf {\bibinfo {volume} {58}},\ \bibinfo {pages} {4395} (\bibinfo {year} {2019})}\BibitemShut {NoStop}%
\bibitem [{\citenamefont {Derickson}(1998)}]{Derickson}%
  \BibitemOpen
  \bibfield  {author} {\bibinfo {author} {\bibfnamefont {D.}~\bibnamefont {Derickson}},\ }\href@noop {} {\emph {\bibinfo {title} {Fiber optic test and measurement}}}\ (\bibinfo  {publisher} {Prentice Hall},\ \bibinfo {year} {1998})\BibitemShut {NoStop}%
\end{thebibliography}%

\end{document}